# QUERYING AND MANIPULATING TEMPORAL DATABASES

Mohamed Mkaouar[1], Rafik Bouaziz[2], and Mohamed Moalla[1]

[1]Université de Tunis El Manar, Faculté des Sciences de Tunis
Campus Universitaire 2092 - El Manar Tunis, Tunisie
Mkaouar.Mohamed@gmail.com, Mohamed.Moalla@fst.rnu.tn
[2]Université de Sfax, Faculté des Sciences Economiques et de Gestion de Sfax
Route de l'Aéroport 3018, Sfax, Tunisie
Raf.Bouaziz@fsegs.rnu.tn

## ABSTRACT

*Many works have focused, for over twenty five years, on the integration of the time dimension in databases (DB). However, the standard SQL3 does not yet allow easy definition, manipulation and querying of temporal DBs. In this paper, we study how we can simplify querying and manipulating temporal facts in SQL3, using a model that integrates time in a native manner. To do this, we propose new keywords and syntax to define different temporal versions for many relational operators and functions used in SQL. It then becomes possible to perform various queries and updates appropriate to temporal facts. We illustrate the use of these proposals on many examples from a real application.*

## 1. INTRODUCTION

Works on temporal databases (temporal DB) [1, 2, 3] aim the modelling and the manipulation of different types of temporal facts, based on two kinds of time: *valid-time* and *transaction-time* [4]. By Timestamping facts with one or both kinds of these times, we distinguish three types of temporal facts: *valid-time facts*, *transaction-time facts* and *bitemporal facts*.

Valid-time facts may relate to the past, to the present or to the future. Each such fact is represented by a timestamp with their valid-times in reality. These times may be instants, time intervals or temporal elements, defined in accordance with a given calendar and a given granularity. Thus, we can maintain the history of all valid facts, which can be updated in real time with retroactive effect or postactive effect. But can not keep track of deletions and corrections of errors.

Transaction-time facts can keep track of the manipulation of facts by the DBMS, which timestamp them by the execution time of the transaction that manipulates the fact. This track covers insert operations, update operations —whether evolution updates or error corrections— and delete operations. Transaction-time facts timestamps are defined according to the schedule adopted by the operating system and with a granularity generally equal to the second, but could be thinner if necessary. Thus, we can maintain the history of all the facts, valid or erroneous, past or current, but not future. Moreover, only the current facts may be updated; the updates can not be made here either with retroactive effect, or with postactive effect.

These valid or transaction histories have then shortcomings. A complete history can be assured only if we timestamp facts by both valid and transaction-times, thus we obtain bitemporal facts.

It then becomes possible to update the facts with retroactive or postactive effects, keep track of valid facts and erroneous facts, and distinguish between valid facts and erroneous facts. However, the management of these facts becomes increasingly complicated and burdensome. Then, the use of temporal dimensions must be justified by the needs of users. A temporal DB must contain conventional facts (non-temporal), when you do not need historical, valid-time facts, when we need a valid history of facts in reality, transaction-time facts, when we need a history that keeps track of the manipulation of facts by the DBMS, and bitemporal facts, when we want to have a complete history.

Despite the large number of works which have been devoted, since the 80s, on temporal DB [1, 2, 3, 4], the current SQL3 standard, updated in 2008, does not yet allow easy definition, manipulation and querying of temporal DBs. Up to the early 2000s, we could understand that the impact of the relational technology had much to do. Indeed, most of the above referred work was carried out by adopting this technology and we know that, in this context, the use of complex data can be made only indirectly by invoking fireworks links, which inevitably impact on the querying and the manipulation. But today, 'Object' technology is growing rapidly, and in this new technology, the use of complex data can be made directly, without the use of particular fireworks. Incorporating the temporal dimension should therefore be simplified. The SQL standard has, in fact, integrated objects concepts since 2003. However, very few studies have focused so far to reconsider the integration of the temporal dimension in this new nail.

This work aims to contribute with concrete proposals along the lines of this technological evolution. Thus, after a general presentation of the state of the art, we begin by recalling the specification of temporal facts with the UML-TF profile [5], then we chain with proposals to a direct and simplified expression of querying and manipulation. A complete example from a real application will serve as a medium of illustration throughout the presentation.

## 2. RELATED WORK

The literature in temporal DB [3] is rich in proposals for querying the different types of temporal facts. Most of these proposals have been made in accordance with the relational technology [6, 7, 8, 9, 10]. However, we believe that the use of the relational model to model temporal DBs complicates the expression of their querying and manipulation. This is already justified by the proposals in the area, through the TSQL2 extension [7]; the language proposed to be integrated to SQL3 with the SQL/Temporal [8] component. Indeed, this component has not been implemented by commercial DBMS and has been suspended from the current version of the standard. The issue has already been raised by other works [11, 12, 13], but the proposed solutions use always the relational technology only.

Following our analysis concerning the syntax, we see that the main 'limit' of the syntax of SQL/Temporal concern the level of his agreement with the used 'Relational' operators and the manner adopted for the evaluation of temporal predicates. Indeed, the proposed new keywords are not targeted to a particular clause to denote a specific operator, but cover the whole query. On the other hand, the evaluation of temporal predicates is based on direct comparisons of the timestamps, using a rich array of features for comparing different types of timestamps, but without using functions of temporal logic that allow to directly evaluate the timestamped columns/tables.

In addition to these theoretical works, possibly accompanied by prototypes [14], there was also some temporal support in commercial DBMS [15, 16, 17]. Among these supports, only Oracle Workspace Manager [17] integrates the valid-time dimension, but the others include the transaction-time dimension and are limited the problem to data recovery. The valid-time support of Oracle is also based on relational technology, allowing only rows (tuples) timestamping. It inspires from SQL/Temporal and is limited to implement functions ensuring the introduction and the comparison of periods. This support has been defined into a comprehensive solution that helps manage collections of updates in separate workspaces from production data, thus it constitutes an ad hoc solution specific to Oracle.

Unlike relational technology, object technology extends to handling different data types (simple, complex, multi-media, semi-structured, etc.). According to this object technology, we note many other works [18, 19, 20, 21, 22]. Among these works, there are extensions to the OQL ODMG standard, proposed in the TOOBIS project [18] and the TEMPO platform [21].

However, unlike SQL, OQL could not succeed in the industrial environment, and we rather experienced the integration of objects concepts to relational products and enjoyed the benefits of both technologies. This is mainly due to the maturity of relational products and the lack of a formal basis for OQL queries. Indeed, these queries are not based on operators of an algebra, but on the object-oriented concepts. Thus, the expression of queries in SQL is simpler and more formal than their expression in OQL. This is also found in the expression of temporal queries in extended OQL. Indeed, in the absence of common operators, temporal extensions have proposed operators that are not easily accepted by non-experts in the field. Also, these extensions, as the base language, do not offer ways to simplify updates in a "non-procedural" manner. As for relational extensions, the proposed temporal extensions to OQL have not been translated into a standard.

Since 2003, the standard database language SQL3 incorporates object concepts in the relational model. However, little Object-Relational temporal extensions have been proposed [23, 24, 25]. In [23] the authors propose a valid-time extension of SQL, based on the model proposed in [11], and its implementation on an Object-Relational DBMS. In [24], the algebra of a generalized temporal database model supporting temporal relations nested to any finite depth is presented; this model deal with the valid-time dimension and the author does not yet, to our knowledge, defines an extension of SQL to support the temporal nested features. The temporal language proposed in [25] integrates only the valid-time dimension at the attribute level, and is also based on SQL/Temporal; we have not perceived a real SQL3 extension, but only how to simplify the expression of temporal predicates with the operators of the temporal logic. Indeed, the language considers that the operators used in this standard —as the cartesian product and the set operators of union, intersection and difference— does not influenced by the introduction of the temporal dimension. However, the intersection of two timestamped columns, for example, must make the intersection of their timestamps. Also, in this language, joins are not treated according to the SQL3 syntax and temporal versions of this operator are not discussed. Finally, we see that the use of **ASELECT** in **SELECT** clause to select timestamped columns is quite complex.

Following this study, we have identified the following objectives:

♦ Relying on a model that allows columns and tables timestamping and supporting the two time dimensions: valid-time and transaction-time. Such a model can be implemented using Object-Relational technology. It allows following the technological evolution, but the data definition language should be further simplified for the definition of BD according to such a model [26].

♦ Propose an SQL3 extension harmonized with all the operators of this language. The elements of the extension should be targeted to the introduction of the temporal dimension, without being connected to other specific aspects.

♦ Simplify the expression of temporal predicates, by maximally avoiding a direct test of timestamps associated with columns or tables.

In the following section, we review on the specification of temporal facts in accordance with the model on which we are based [5, 26], and we continue in the other two sections to present elements of the extension on querying and manipulation.

## 3. MODELLING TEMPORAL DATABASES WITH UML-TF

UML-TF is an UML profile dedicated to the modelling of temporal facts [5]. With this profile, modelling is made through the various levels of abstraction proposed by the MDA (Model Driven Architecture) approach, while expressing the temporal dimensions of any facts by stereotypes that we classify into three categories: valid-time stereotypes, transaction-time stereotypes and bitemporal stereotypes. To these three categories we associate the following icons:

♦ '' which symbolizes the real world clock, and is intended to the first category of stereotypes. This icon is used at the CIM (Computational Independent Model) level according to the MDA approach.

♦ '' which represents the clock of the machine, and is intended to the second category of stereotypes. This icon is used in the PIM (Platform Specific Model) level of MDA.

♦ '' which symbolizes both clocks, and is intended to bitemporal stereotypes. Such icon results from any two declarations, first by a valid-time icon, then by a transaction-time icon.

The use of the proposed stereotypes is done in UML-TF by placing the icon before the name of the concerned element. To realize the UML-TF profile, we defined an abstract stereotype for each of the three categories of stereotypes. Each abstract stereotype is characterized by meta-properties to be able to express, for each stereotyped element, timestamps properties as well as constraints on these timestamps or on the modelled temporal instances [26]. Further details concerning the definition of the UML-TF stereotypes can be found in [5].

The class diagram of Figure 1 illustrates an UML-TF model. It is about an application that manages information on people, each of which is characterized by a number, a last name, a first name, a gender, a birth date, a nationality, a home address, an email address and telephones numbers. A person, whether student or teacher or both, may exercise more than three activities. A student can become a teacher, while keeping his student status. We must know in addition the high school diploma obtaining date and possible diseases of each student, and the status, grade and salary of each teacher.

A student is assigned during a period to a group of a given audience. A group is characterized by a code and a number of students, and an audience is characterized by a code, a label and a number of groups. Both groups that audiences are to historicize as they evolve in reality.

A teacher belongs to a department and may be responsible for more than nine modules. A department is described by a code, a label and a budget, and is headed by a leader among teachers. A module is described by a number and a designation and can be taught to several audiences, each with a given period and a coefficient. Between seven and fifteen modules are taught to any given audience. Any teaching session (education) concerns a teacher, a module, a group of an audience or an audience; it is described by a type (Course, TD, TP), a day, an hour and a classroom number.

Each student should have, for each module that he studies, a test mark, a mark for the principal exam, and possibly a mark for the second exam.

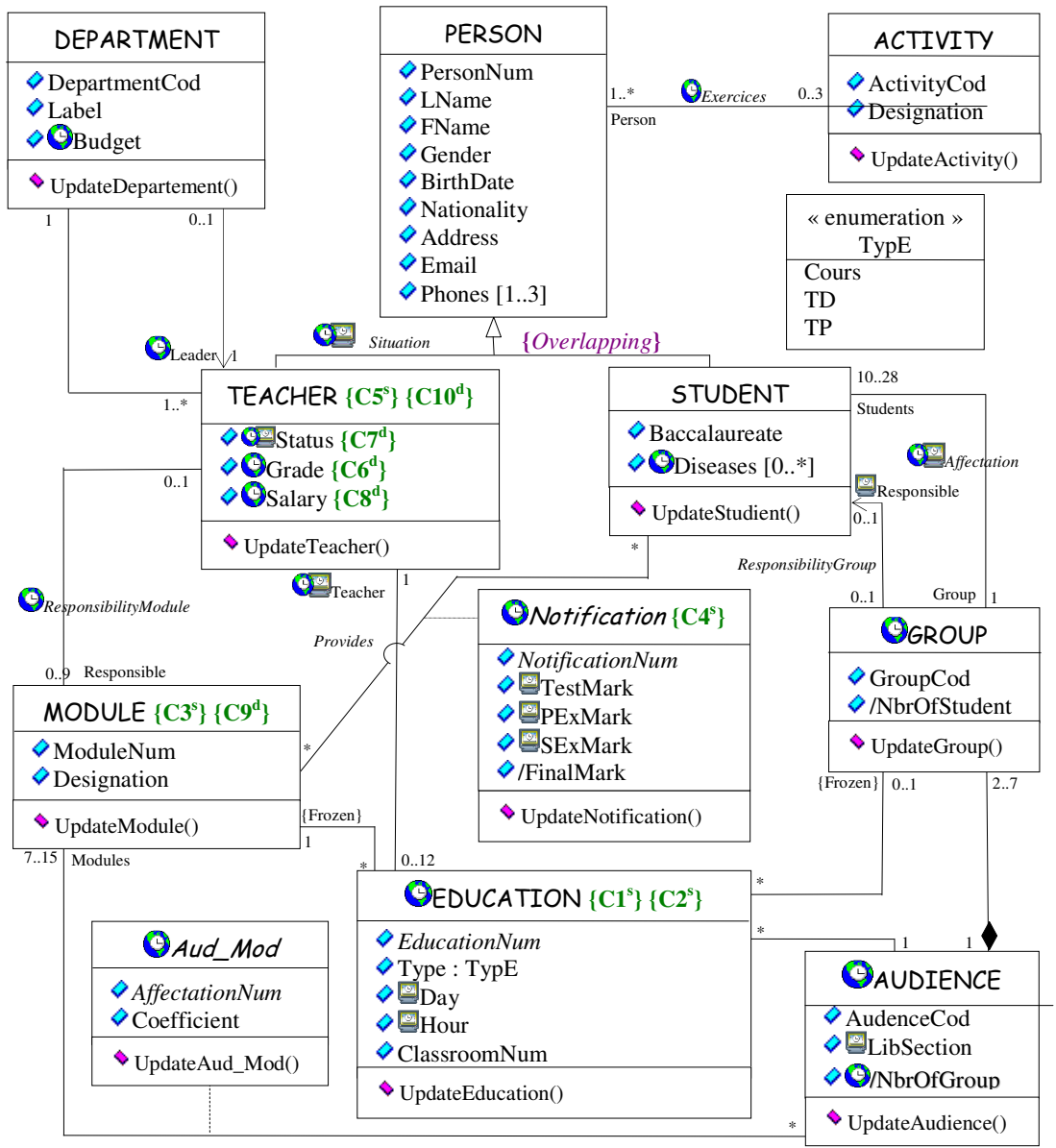

Figure 1. UML-TF class diagram modeling the database of an application for a higher education institution.

A more detailed description of this application, especially the constraints represented in the diagram ({**C1$^s$**} ... {**C10$^d$**}), is found in [26]. The Object-Relational schema resulting from the mapping of this class diagram is as follows:

**AUDIENCE V**      (<u>AudienceCod</u>, LibSection, NbrGroup **V**)

**GROUPE V**        (<u>GrpoupCod</u>, NbrStudent, StudentResponsibleNum # **T**, AudienceCod#)

**ACTIVITY**        (<u>ActivityCod</u>, Designation)

**PERSON V T**      (<u>PersonNum</u>, LName, FName, Gender, BirthDate, Nationality, Adress, Email, [Phones][3], [ActivitityCod **V**][3])

**STUDENT**         **INHERITENCE FROM PERSON** (Baccalaureate, [Diseases **V**][*], GroupCod# **V  T**)

**TEACHER**      **INHERITENCE FROM PERSON** (Status **V**  **T**, Grade **V**, Salary **V**, DepartementCod#)

**DEPARTEMENT**   (<u>DepartementCod</u>, Label, Budget **V**, TeacherLeaderNum# **V**)
**MODULE**        (<u>ModuleNum</u>, Designation, TeacherResponsibleNum# **V**)
**AUDIT_MOD V**  (<u>AffectationNum</u>, ModuleNum#, AudienceCod#, Coefficient)
**EDUCATION V**   (<u>EducationNum</u>, TeacherNum# **V T**, ModuleNum#, AudienceCod#, GroupCod#, Type, Day **T**, Hour **T**, ClassroomNum)
**NOTIFICATION V** (<u>NotificationNum</u>, StudentNum#, ModuleNum#, TestMark **T**, PExMark **T**, SExMark **T**, FinalMark)

In the above Object-Relational schema, a letter 'V' indicates a valid timestamp, a letter 'T' indicates a transaction timestamp, underlined columns indicates primary keys, a column followed by '#' symbol indicates a foreign key, and the multiplicities of the multi-valued columns are indicated in superscripts between brackets.

On the other hand, every timestamped column is a multi-valued column because it can have multiple values, each of which is associated with one/two timestamp(s). This structure characterizes an Object-Relational schema from a relational schema, and thus avoids the use of a separate table and a relationship between this table and the parent table for any temporal column. As an example, we present in Table 1 an excerpt from an extension of the TEACHER table. For readability, we do not present transaction-time values which are temporal values with a granularity of seconds.

Table 1.  An excerpt of an extension of the TEACHER table.

| TEACHER | TeacherNum | ... | Status | | | Grade | | ... | DCod | Validity |
|---|---|---|---|---|---|---|---|---|---|---|
| | | | **Value** | **Validity** | **Transaction** | **Valus** | **Validity** | | | |
| | 777 | ... | 'R' | ['2008'-'uc') | ['T₅'-'now') | 'Pr' | ['2010'-'uc') | ... | CS | ['2002'-'uc') |
| | | | 'PS' | ['2005'-'2008') | ['T₃'-'T₄') | 'AP' | ['2004'-'2010') | | | |
| | | | 'T' | ['2002'-'2005') | ['T₁'-'T₂') | | | | | |
| | 555 | ... | 'PS' | ['2010'-'uc') | ['T₆'-'now') | 'AP' | ['2010'-'uc') | ... | CS | ['2010'-'uc') |

***Legend :*** R' : Recruited ; 'PS' : Permanent by Stage ; 'T' : Temporary ; 'AP' : Associate
Professor ; 'Pr' : Professor ; 'CS' : Computer Science.

## 3. QUERYING TEMPORAL DATABASES

The enrichment that we propose for querying temporal DBs is mainly based on seven temporal 'terms' expressed by new keywords that can distinguish seven temporal versions of the relational operators and those specific to SQL. These terms may be used in different clauses of a query. This enrichment also includes means for temporal grouping.

By default, the proposed terms concerning the valid-time dimension of the used column/table; however, we can express this explicitly by writing the letter 'V' before the used term. Some of them can also be applied to the transaction-time dimension, this must be stated explicitly by mentioning the letter 'T'.

Recall that when timestamping with only valid-times, we can maintain the history of all the valid facts, without being able to keep track of deletions and corrections of errors. When timestamping with only transaction-times, we can keep track of both valid and erroneous facts, but we can not distinguish between these two fact types. Only when timestamping with the two kinds of times, we can maintain and distinguish these two types of facts, and we can distinguish, in addition, the updates made with retroactive effect and these performed with postactive effects.

In the following, we study the new meanings of the SQL operators and functions where they are applied with temporal columns or temporal tables and in the presence of one of the proposed terms, by considering a single temporal dimension, *i.e.*, the valid-time dimension or the transaction-time one. At this level of application, the user must know what the meaning of the timestamped data to be queried, especially when he concerned by the transaction-time dimension with transaction-time facts or with bitemporals facts. We propose in §4.1.6 elements which consider the two temporal dimensions and which apply only to bitemporal facts.

## 4.1. Temporal versions of relational operators

We define seven terms to integrate temporal specifications to the relational operators (restriction, natural join, negation, division, cross product and set operators). These terms are expressed by the following keywords: **HISTORY**, **PAST**, **FUTURE**, **@** 'Dat', **BETWEEN** 'Dat1' **AND** 'Dat2', **WHEN** 'Condition', and **{BEFORE | AFTER | SINCE}** **{**'Dat' | 'Condition'**}**. Each of these keywords sets the considered operator by a date or time period.

All these seven terms apply to the valid-time dimension for all operators. **FUTURE** can not apply with the transaction-time dimension. The other six terms apply to this dimension only for the restriction operator.

### 4.1.1. Temporal restriction (selection)

This temporal operator applies on temporal columns or temporal tables, and allows for a restriction on values or rows. Table 2 details the meaning of each of the proposed temporal terms when applied with this operator.

Table 2. Meaning of the restriction of temporal columns or temporal tables.

| | |
|---|---|
| **HISTORY** {Column \| Table} | Indicates all values/rows, with their timestamps. |
| **PAST** {Column \| Table} | Indicates all past values/rows, with their timestamps. |
| **FUTURE** {Column \| Table} | Indicates all valid values/rows in the future, with their timestamps. |
| {Column \| Table} **@** 'Dat' | Indicates all values/rows in the specified date. |
| {Column \| Table} **BETWEEN** 'Dat1' **AND** 'Dat2' | Indicates values/rows and their timestamps, corresponding to the specified period. |
| {Column \| Table} **WHEN** 'Condition' | Indicates values/rows and their timestamps during the period when the condition is verified. |
| {Column \| Table} **{SINCE \| BEFORE \| AFTER}** {'Dat' \| 'Condition'} | Indicates values/rows and their timestamps during the mentioned period, being able to be since, before or after the date 'Dat', or since, before or after the check of the condition 'Condition'. |

***Examples:*** According to the Object-Relational schema corresponding to the class diagram in Figure 1, we give below some applications of temporal restrictions:

♦ **FROM** STUDENT: current students.

♦ **FROM** **HISTORY** STUDENT: all valid rows (past, current and possibly future) of students, considering the valid timestamps associated with each row.

♦ **FROM** STUDENTT **BETWEEN** '2002' **AND** '2006' **T**: valid and erroneous rows of students at the indicated period considering the transaction-time timestamps associated with each of them. This expression can be written differently as follows: STUDENT **[**'2002' **-** '2006'**] T**.

♦ **FROM** TEACHER **WHEN** LName = 'ABDELWAHEB' **AND** FName = 'Mohamed' **AND** Grade = 'Professor': valid rows of teachers during the period when Mohamed ABDELWAHEB is Professor.

♦ **FROM** TEACHER **[SELECT** E.**V FROM** STUDENT **WHERE** LName = 'TOUNSI' **AND** FName = 'Feres']: valid rows of teachers in the period of the studies of Fères TOUNSI.

To improve performance, the execution of these restrictions is to perform before the execution of joins of the considered query.

Specifically for restrictions we propose, beside the seven temporal terms detailed in Table 2, new functions involving restrictions on temporal columns. These functions are presented in Table 3; the first column shows the function name, the second column gives a brief description, and the third column states, using the symbol ⌷ ', if the function can be applied by considering the transaction-time dimension.

Table 3.  Proposed functions to extract values/timestamps of temporal columns.

| | | |
|---|---|---|
| {FIRST | LAST} Column | Indicates the first/last value of the column and its timestamp. | Tt |
| PREVIOUS ['Val'] Column | Indicates the value which precedes the current value, or the value 'Val', and its timestamp. | Tt |
| NEXT ['Val'] Column | Indicates the valid value which follows the current value, or the value 'Val', and the corresponding valid-time timestamp. | ☐ |
| PREVIOUS_SCALE Column | Indicates the value which precedes the current value and its timestamp, according to the specified granule. | Tt |
| NEXT_SCALE Column | Indicates the valid value which follows the current value and its timestamp, according to the specified granule. | ☐ |
| EVOLUTION HISTORY Column | Indicates all the evolution dates of the column. | ☐ |
| EVOLUTION Column | Indicates the evolution date to the current value. | ☐ |
| {FIRST | LAST} EVOLUTION Col | Indicates the first/last evolution date of the column. | ☐ |
| EVOLUTION 'Val1'-'Val2' Col | Indicates the evolution date from 'Val1' to 'Val2'. | ☐ |
| Column TIMESTAMPS 'Val' | Indicates the timestamp associated to the value 'Val'. | Tt |
| THIS Column.Value | Indicates the value associated with a date specified somewhere in the query. | Tt |
| THIS Column.Timestamp_Name | Indicates the timestamp associated with a value specified somewhere in the query. | Tt |

The restrictions on temporal columns and the functions shown in Table 3 can be used both in the **SELECT** and in the **WHERE** clauses. In the latter case, they can simplify the expression of temporal predicates. We propose to define, on the same principal, other functions to more simplify the expression of these predicates, as: **ALWAYS** Column, **ANYTIME PAST** Column, **INCREASE** Column, **FIRST INCREASE** Column, and **DECREASE** Column.

***Examples:*** considering a given teacher, we illustrate the use of temporal restrictions in the **SELECT** clause, as follows:

♦ **SELECT** Status: his actual status.

♦ **SELECT** FIRST Status: his first status, also by showing the corresponding valid timestamp.

- ◆ **SELECT** LAST Status **S,** THIS **S.**T: his last status, also by showing the two corresponding timestamps.

- ◆ **SELECT** Status @ '2/1/2008': his valid status on '2/1/2008'.

- ◆ **SELECT** HISTORY **T** Status: all his statuses with the corresponding transaction timestamps.

- ◆ **SELECT** PAST | FUTURE Status: all his valid statuses in the past/future; with each status we also show the corresponding valid timestamp.

- ◆ **SELECT** Status WHEN Initcap(Grade) = 'Assistant Professor': all his valid status when he is Assistant Professor.

- ◆ **SELECT** Grade SINCE Initcap(Status) = 'Recruited': all his valid grades since his evolution to the 'Recruited' status.

- ◆ **SELECT** EVOLUTION Status: the evolution date of the current status.

- ◆ **SELECT** LAST EVOLUTION Status: the last evolution date of his status.

- ◆ **SELECT** EVOLUTION 'Contractual'-'Permanent by Stage' Status: the evolution date of his status from 'Contractual' to 'Permanent by Stage'.

- ◆ **SELECT** Status.V TIMESTAMPS 'Contractual': the valid timestamp associated to the 'Contractual' value of his status.

- ◆ **SELECT** TEACHER.T: the transaction timestamp of a teacher.

- ◆ **SELECT** THIS Statut.V: the valid timestamp of a status value indicated somewhere in the query.

Let us illustrate now some applications in the **WHERE** clause:

- ◆ **WHERE** 'Swimming' **IN** HISTORY(Initcap(Activities)): students/teachers whose exercised swimming.

- ◆ **WHERE** 'Assistant Professor' **IN** HISTORY(Initcap(Grade G)) **AND** THIS G.V LonguestThen 7 YEARS: teachers who remained with the 'Assistant Professor' grade during a period more than 7 years.

- ◆ **WHERE** EVOLUTION Grade BEFORE (CurrentDate - 5 YEARS): teachers who have not changed a grade since 5 years.

- ◆ **WHERE** BEGIN(FIRST(Grade.V)) = '2005' **AND** THIS Grade = 'Assistant Professor': teachers recruited with the 'Assistant Professor' grade in 2005.

- ◆ **WHERE** ALWAYS Budget >= 300: departments which never had a budget lower than 300.

- ◆ **WHERE** ANYTIME PAST Budget < 300: departments which had, in the past, a budget lower than 300.

- ◆ **WHERE** ALWAYS Budget >= 300 WHEN TeacherLeaderNum = 555: departments which did not have a budget lower than 300 during the period of their direction by the teacher identified by 555.

### 4.1.2. Temporal natural join

This temporal operator is to apply when we have a temporal relationship between two tables that can be temporal or non-temporal; such relationship is realized by a temporal column and a referential constraint. The verification of the timestamps of these columns, with respect to the possible timestamps associated with the concerned objects, is made at the introduction of the links of the considered relationship, using the constraints declared at the creation of the DB [26]. The temporal join operator is then dependent only on the timestamps of the links. Table 4 details the meanings of the seven proposed versions of the temporal join operator.

Table 4. Proposed versions of temporal join between tables.

| HISTORY *Natural Join* | Considers all valid links, as well as current, past and future links. |
| PAST *Natural Join* | Considers only valid links in the past. |
| FUTURE *Natural Join* | Considers only valid links in the future. |
| *Natural Join* @ 'Dat' | Considers valid links at the specified date. |
| *Natural Join* BETWEEN 'Dat1' AND 'Dat2' | Considers valid links during the specified period. |
| *Natural Join* WHEN 'Condition' | Considers valid links during the period when the condition is verified. |
| *Natural Join* {SINCE \| BEFORE \| AFTER} {'Dat' \| 'Condition'} | Considers valid links during the mentioned period: since, before or after the date 'Dat', or since, before or after the verification of the condition 'Condition'. |

In the Object-Relational model, joins can also be made by objects references; from a reference we may have access to some values of an object using the dot notation. This notation should also be extended to consider the various versions of temporal joins.

***Examples*** of temporal join in the **FROM** clause:

♦ **FROM** TEACHER **Natural Join** DEPARTMENT: current links between the two tables; each teacher is linked with his department.

♦ **FROM** TEACHER **Natural Join [**'1997'-'2003'**]** DEPARTMENT: links between these two tables, valid during the period overlapped with the mentioned period; each link is considered with the overlapping period.

♦ **FROM** TEACHER **Natural Join [SELECT** E.**V FROM** STUDENT **WHERE** LNom **=** 'ABDELWAHEB' **AND** FName **=** 'Mohamed' **AND** Grade **=** 'Professor'**]** DEPARTMENT: links between these two tables, valid during the period overlapped with the period when Mohamed ABDELWAHEB is Professor.

### 4.1.3. Temporal negation and temporal division

Negations and divisions are expressed in SQL with correlated blocks, using the **NOT EXISTS** operator and joins between the used tables. In a similar way to what is done for the temporal joins, temporal versions of negation and division use temporal links between involved tables. These temporal operations are also enriched by checking one of the conditions that we indicated for each version of temporal join, considering only links timestamps.

***Example:*** To know the groups never having students of French nationality, the following query is written:

```
SELECT        *
    FROM        HISTORY GROUP G
    WHERE        NOT EXISTS  (SELECT        *
                            FROM        HISTORY STUDENT S
                            WHERE        Nationality = 'French'
                                    AND G.GroupeCod IN HISTORY (S.GroupeCod) );
```

***Example:*** To know the current teachers who taught or teach all the current audiences, the following query is written:

```
SELECT      *
   FROM   TEACHER T
   WHERE  NOT EXISTS  (SELECT      *
                       FROM      AUDIENCE A
                       WHERE     NOT EXISTS  (SELECT      *
                                             FROM      HISTORY EDUCATION E
                                             WHERE     E.TeacherNum = T.TeacherNum
                                             AND E.AudienceCod = A.AudienceCod));
```

Note that the expression of these operators in SQL is not simple. To this end, we think that it is useful to enrich SQL with new keywords to simplify the expression of those operators. With these enrichments, the proposed temporal versions of these operators become much simpler. We plan to address this aspect in a future paper.

### 4.1.4. Temporal cartesian product (cross product)

This temporal operator is applied between two temporal tables. Contrary to a temporal join iteration, which considers a single timestamp (that of the link), a temporal cartesian product iteration considers two timestamps (those of the involved rows). The temporal cartesian product operation consists to make the conventional cartesian product between the two tables, while associating with each concatenation one timestamp that corresponds to the intersection of the two timestamps of the concerned rows. We detail versions that we propose for this operator in Table 5.

Table 5. Temporal versions of the cartesian product between tables.

| | |
|---|---|
| HISTORY ***Cross Join*** | Considers all concatenation between all valid rows, as well as current, past or future. |
| PAST ***Cross Join*** | Considers only concatenation between past valid rows. |
| FUTURE ***Cross Join*** | Considers only concatenation between future valid rows. |
| ***Cross Join*** @ 'Dat' | Considers concatenation between rows valid at the specified date. |
| ***Cross Join*** BETWEEN 'Dat1' AND 'Dat2' | Considers concatenation between rows valid during the specified period. |
| ***Cross Join*** WHEN 'Condition' | Considers concatenation between rows valid during the period when the condition is verified. |
| ***Cross Join*** {SINCE \| BEFORE \| AFTER} {'Dat' \| 'Condition'} | Considers concatenation between rows valid during the mentioned period: since, before or after the date 'Dat', or since, before or after the verification of the condition 'Condition'. |

### 4.1.5. Temporal set operators

Similarly to non-temporal set operators, temporal set operators —temporal set intersection, temporal set union and temporal set difference— operate on the results of two sub-queries to perform the intersection, the union or the difference between the values/rows of these two sub-queries. A temporal intersection proceeds, in addition, to the intersection of the timestamps of the considered common values/rows. Concerning a temporal union (or temporal difference), it proceeds to the union (or difference) of the timestamps of the common values/rows.

*Example:* The common number of groups between the audience 'Marketing' and the audience 'Management':

**SELECT** HISTORY(NbrGroup) **FROM** AUDIENCE **WHERE** LibSection **=** 'Marketing'
**INTERSECT**
**SELECT** HISTORY(NbrGroup) **FROM** AUDIENCE **WHERE** LibSection **=** 'Management'**;**

### 4.1.6. Bitemporal versions of SQL operators

With the presence of the two time dimensions for a column or a table, it is possible to find specific information, such as data introduced with retroactive effects, with postactive effects or erroneous data. To simplify the search of these data, we propose to define bitemporal versions for each of the operators and functions of SQL, with the exception of the cross product operator. These versions are to express in the same principle, with appropriate keywords, namely **RETROACTIF** [**WITH** Value SCALE], **POSTACTIF** [**WITH** Value SCALE] and **ERRONEOUS**. The option '**WITH** Value SCALE' can specify a phase period, from the moment of the execution of the transaction inserting the data.

*Examples:* We illustrate some applications of these bitemporal versions in the two clauses, **SELECT** and **FROM**, as follows:

♦ **SELECT** **RETROACTIF | POSTACIF | ERRONEOUS** Status: all statutes introduced with retroactive effect, with postactive effect or the erroneous values introduced into the statutes of one or several teachers.

♦ **FROM** TEACHER **POSTACTIF** **Natural Join** DEPARTMENT: natural join involving links defined with a postactive effect between these two tables.

## 4.2. Temporal extensions to the specific operators and functions used in SQL

Recall that SQL uses the conventional operators for manipulating numeric values, string functions and logical operators. It also uses aggregate functions —**Count()**, **Sum()**, **Min()**, **Max()** and **Avg()**— for performing calculations on a set (or groups) of rows.

In the absence of one of the keywords defined in §3.1.1, these operators and functions are interpreted as before, by taking into consideration only the current values/rows. Otherwise, a temporal version is needed for each of these operators and each of these functions, which is not to represent explicitly. We briefly detail the versions that we propose in the following two sub-sections.

### 4.2.1. Temporal version for conventional operator

A temporal version of a conventional operator provides a set of values for each identified object. Each value corresponds to a common timestamp between the two operands.

*Example:* The following query returns the result of the addition of the marks of the student N° 0900105 in the module N° 25 for each period. In the absence of the HISTORY keyword, this query gives the one possible value of the current period.

**SELECT** TestMark **+** PExMark
    **FROM** HISTORY NOTIFICATION
    **WHERE** StudentNum **=** 0900105 **AND** ModuleNum **=** 25**;**

### 4.2.2. Temporal version for aggregate functions

The temporal version of an aggregate function considers not only the current rows, but also past and future ones that verifying the predicate of the query. This is the same case if the query uses or not **GROUP BY**.

***Examples:*** we illustrate the use of temporal version for the aggregate functions as follows:

♦ **SELECT** Max(HISTORY Budget): extract the maximum budget among all the values from the budgets (current, past and future) of the considered department(s).

♦ **SELECT** Sum(HISTORY Budget): extract the sum from the budgets of the considered department(s).

♦ **SELECT** Max(HISTORY Budget DECADE): extract the maximum budget among all the values from budgets (current, past and future) of the considered department(s), in each decade.

The use of the proposed temporal versions for the aggregate functions must be performed in an appropriate manner to well target the values to be account for the set of the considered rows or each group of rows.

### 4.3. Temporal grouping of values and rows

The temporal grouping that we propose operates on values of a temporal column or rows of a temporal table, according to a temporal granule. By using the granules of the Gregorian calendar, it is possible to group these values or rows in each decade, each year, each semester, etc.

A temporal grouping on values of a temporal column is indicating by expressing the granule behind the name of the considered column. A temporal grouping on rows is expressed behind the name of a temporal table or in the **GROUP BY** clause. In the latter case, this grouping can be used with or without one/several other(s) column(s) of the table.

***Examples:*** We illustrate the use of the temporal grouping as follows:

♦ **SELECT** Budget DECADE: grouping of the budget, of one or several departments, in each decade.

♦ **GROUP BY** DepartmentCod, YEAR: grouping employees by department and year.

We also propose to allow the 'filtering' of a temporal grouping, by conducting tests on the results returned by an aggregate function applied to temporal groups. To be compatible with SQL, this filtering is always expressed in the **HAVING** clause, without defining a specific other clause.

## 3. INSERTION, MODIFICATION AND DELETION OF TEMPORAL FACTS

With the consideration of temporal facts in DBs, the three manipulation commands require deep enhancements. Indeed, (i) the **INSERT** command should allow the introduction of valid timestamps and the allocation of transaction timestamps, and (ii) the **UPDATE** command must carry out various "forms" of non-destructive updates and (iii) the **DELETE** command must allow the non-destructive delete of rows, using transaction timestamps. We further detail the enhancements that we propose in the following three sub-sections.

## 5.1. Insertion of temporal rows

The **INSERT** command should allow the user to introduce valid timestamps for any valid-time or bitemporal value/row. If such timestamps are not specified, default values, referred at the creation of the DB, are used. Also, this command has to associate transaction timestamps to any transaction-time or bitemporal value/row. These timestamps are to extract from the schedule of the machine, at the validation of the concerned transaction.

*Examples:*

♦ The insertion of a row in the TEACHER table is done as follows:

> **INSERT INTO** TEACHER **VALUES** (0900105, 'TOUNSI', 'Fères', 'M', '1/5/1973', 'Tunisian', 'Tunis', Null, {}, {'Tennis' ['2003'-'2004'] **U** ['2010'-'uc'], 'Swimming' ['2009'-'uc']}, {'Recruited' ['2010'-'uc']}, {'Assistant Professor' ['2009'-'uc']}) ['1997'-'uc'];

♦ The insertion of two rows in the MODULE table is done as follows:

> **INSERT INTO** MODULE **VALUES** (25, 'DATABASES', {0355 ['2010'-'uc']}), (15, 'MARKETING', {0502 ['2010'-'uc']});

## 5.2. Modification of values

To be able to perform various "forms" of non-destructive update of columns, we propose to distinguish these forms in the **UPDATE** command with three options: **SET**, **TAG ON** and **CORRECT**. The application of these options for single-valued columns is described below.

The **SET** option allows you to change column values that are not associated with valid timestamps. When a column is non-temporal, the change is to achieve in a destructive manner, the new value replaces the old one. But, when a column is timestamped with a transaction-time, the change is to achieve in a non-destructive manner, the old value is first stored after timestamping it by the end transaction-time, then the new value, timestamped by the begin transaction-time, is assigned to the column.

*Example:* The modification of one or several nationalities is done as follows: **SET** Nationality = 'French'.

The **TAG ON** option allows the introduction of new values for columns that are associated with valid timestamps. This option allows the evolution of values according to the reality, the introduction of missing past values, and the introduction of future values in advance. Each value is indexed by the corresponding valid timestamp. The introduction of a new value is rejected if there is already another value for the same column with a valid timestamp that overlaps the timestamp of the value to be introduced.

*Example:* The addition of values for the grade and the status is done as follows: **TAG ON** Grade 'Assistant Professor' ['2010'-'UC'], Status {'Temporary' ['1997'-'1998'], 'Contractual' ['1999'-'2005'], 'Permanent by Stage' ['2006'-'2010']}.

The **CORRECT** option allows the correction of columns which are associated with valid timestamps. With this option, we can correct a value, its valid timestamp, or both at once. Corrections of timestamps associated with rows are also performed by this option. In all cases, corrections are realized in a destructive manner to the valid-time columns/rows and in a non-destructive manner to the bitemporal columns/rows.

***Example:*** The correction of the assignment of a teacher to a teaching session (education) is done as follows: **CORRECT** TeacherNum = **(SELECT** TeacherNum **FROM** TEACHER **WHERE** Upper(Nom) **= '**TOUNSI**' AND** Initcap(Prenom) **= '**Fères'**)** ['2010'-'2011'].

For both options **SET** and **CORRECT**, it is possible to use the missing value "Null" to delete —in a destructive or non-destructive manner— a current value that is not associated with a valid timestamp, or to delete —in a destructive or non-destructive manner— a valid value that is associated with a valid timestamp, respectively.

### 5.3. Deletion of rows

With the introduction of the time dimension in DBs, there are destructive deletions and non-destructive ones. Delete a row to which is not associated a transaction timestamp (either to the row, or to a value of one of its columns) is to realize in a destructive manner. But the deletion of a row to which is associated a transaction timestamp is to realize in a non-destructive manner.

However, in the second case, rows can be destructively (physically) deleted according to the vacuuming parameters, when these parameters (maximum lifecycle/number of past values or rows to keep) have been defined for the concerned table [26]. To allow the destructive deletion of such rows without considering these parameters, we propose to introduce a new command, **VACUUM**. The syntax of the use of this command is similar to the **DELETE** command, but their privileges should be granted in a strict manner.

In any case, it is not possible to delete a "master" row attached to "detail" rows. The purpose of this treatment is to indicate at the creation of the DB with the referential integrity parameters, via the '**ON DELETE CASCADE**' or '**ON DELETE CORRECT** Null' options.

## 6. CONCLUSIONS AND FUTURE WORK

In this paper, we have studied how we can simplify querying and manipulating temporal DBs. Based on a model that integrates time in a native manner [5, 26], we then have proposed temporal extension to the standard SQL3 by new keywords. These keywords are used to define several versions of temporal operators to SQL, and then to perform many "forms" of querying and updating temporal DBs. As is illustrated by different examples from a real application, the proposed enhancements simplify the expression of temporal queries and temporal manipulations.

In the future, the formal semantics of our proposals will be proposed. We also plan to implement these proposals, while exploring new techniques for indexing and presentation. This implementation must ensure upward compatibility with the standard and with the existing applications [27]. Thus, it is guaranteed that legacy queries results remain the same when these queries run on DBs incorporating the temporal dimension, and they always bring current data.